\documentclass{elsart}

\usepackage{epsfig}

\usepackage{amssymb}

\begin{document}

\begin{frontmatter}


\title{Price fluctuations from the order book perspective
- empirical facts and a simple model.}


\author{Sergei Maslov}

\address{Department of Physics, Brookhaven National Laboratory, 
Upton, NY 11973, USA\thanksref{email_sm}}
\thanks[email_sm]{E-mail: maslov@bnl.gov}

\author{Mark Mills}
\address{1320 Prudential, Suite 102, Dallas, TX 75231, USA\thanksref{email_mm}}
\thanks[email_mm]{E-mail: mmills@htcomp.net}

\begin{abstract}
Statistical properties of an order book and the effect they have on price dynamics were 
studied using the high-frequency NASDAQ Level II data. It was observed that
the size distribution of marketable orders (transaction sizes) has power
law tails with an exponent $1+\mu_{market}=2.4
\pm 0.1$. The distribution of limit order sizes  
was found to be consistent with a power law with an exponent close to 2. 
A somewhat better fit to this distribution 
was obtained by using a log-normal distribution with an effective power law 
exponent equal to 2 in the middle of the observed range. The depth of
the order book measured as a price impact of a hypothetical large
market order was observed to be a non-linear 
function of its size.
A large imbalance in the  number of limit orders placed at bid and ask sides 
of the book was shown to lead to a short term deterministic price
change, which is in accord with the law of supply and demand. 
\end{abstract}

\end{frontmatter}

As a result of collective efforts by many authors the list of basic
``stylized'' empirical facts about market price fluctuations has now 
begun to emerge \cite{thisvolume}.
It became known that the histogram of short term price
fluctuations $\delta p(t)=p(t+\delta t)-p(t)$ 
has ``fat'' power-law tails: ${\rm Prob}(\delta p>x) \sim x^{-\alpha}$.
The exponent $\alpha$ was measured to be close to 3
in major US markets \cite{gopi1} as well as foreign exchange markets
\cite{dacorogna}. The other well established empirical fact is
that while the sign of $\delta p(t)$ measured at different times has
only short term correlations, its magnitude $|\delta p(t)|$ (or
alternatively its square $\delta p(t)^2$) has a long term memory as 
manifested by slowly decaying correlations. The correlation function 
was successfully fitted by a power law $t^{-\gamma}$ 
with a small exponent $\gamma \simeq 0.3$ \cite{cont,liu} 
over a rather broad range of times. 

Several simplified
market models were introduced in an attempt to reproduce and explain 
this set of empirical facts \cite{cmz,cont_bouchaud}.
The current consensus among econophysicists seems to be that these facts 
are a manifestation of some kind of strategy herding effect,
in which many traders lock into the same pattern of behavior. 
Large price fluctuations are then explained as a market impact 
of this coherent collective trading behavior. Any model aiming 
at understanding price fluctuations needs to
define a mechanism for the formation of the price. Here the usual
approach is to postulate some empirical (linear or non-linear) 
market impact function, which reduces calculating prices 
to knowing the imbalance between the supply of and the  
demand for the stock at any given time step. 

Recently one of us (SM) has introduced a toy model 
\cite{maslov} in which the same standard set of 
stylized facts, albeit with somewhat 
different critical exponents, was generated in 
the absence of any strategic
behavior on the part of traders. The model uses a 
rather realistic order-book-based mechanism of price formation, 
which does not rely on any  postulated 
market impact function. Instead, price fluctuations 
arise naturally as a result of changes in the balance of orders in the
order book. The long memory of individual entries in this book gives rise to 
fat-tailed price distributions and volatility clustering. 
Every market has two basic types of orders, which we would refer to
as limit and market orders.  A limit order to sell (buy) is an
instruction to sell (buy) a specified number of shares
of a given stock if its price rises above (falls below) a predefined
level, which is known as the execution price of a limit order. A
market order on the other hand is an instruction to immediately sell (buy)
a specified number of shares at whatever price currently available at
the market. Here we do not make a distinction between a true market
order and a marketable limit order, placed at the 
inside bid or ask price, and refer to both of them as
'market orders'. The model of Ref. \cite{maslov} assumes the simplest
possible mechanism for the dynamics of individual 
orders in the order book. At each step a new order is submitted to the market. 
With equal probabilities this order can be a limit order to sell, a market
order to sell, a limit order to buy, or a market order to buy. 
All orders are of the same unit size, and a new limit order to 
sell (buy) is placed with a random offset $\Delta$ above (below) 
the most recent transaction price. In spite of its utmost simplicity
the  model has a surprisingly rich behavior, which up to now was
understood only numerically. The distribution of price fluctuations has
power law tails characterized by an exponent $\alpha=2$, while the
correlation function of absolute values of price increments decays
as $t^{-0.5}$. 

Of course, the dynamics of a real order book is much more complicated
than rules of the toy model from Ref. \cite{maslov}.  
First of all, in real markets, both market and limit orders come in vastly 
different sizes and exist for various time frames.
Secondly, participants of real markets do use strategies after all.
In particular, both under-capitalized speculators and well-capitalized
market makers avoid static public display of their willingness to accept a given 
price, and adjust their limit order size and price regularly.
Finally, there is a practically all-important matter of time delay between 
the actual state of the order book and whatever a particular trader 
observes on his/her screen. Prior to electronic 
data transmission, investors might not know at what price 
the queue is matching their buy and sell orders until long after 
the transaction took place.  On the other hand, market makers have always 
had near immediate access to completed transaction data.  
With modern computerized markets, there is a much shorter 
delay between a transaction's completion time and trader's 
awareness of the event, but the delay still 
exists. The inhomogeneity of those delay times for different 
market participants contributes to the wide variety of strategies 
employed by traders.

In this work, we attempt to establish some empirical facts about the 
statistical properties and dynamics of publicly displayed limit 
orders using data collected in a real market.  The purpose of 
this analysis is twofold.  First of all, these new observations 
would extend a rather narrow list of stylized facts about real 
markets.  As in other branches of physics (or any other empirical 
science for that matter) the only way to choose among many 
competing theoretical models is to make new empirical 
observations.  Since the high frequency data about the state 
of an order book is much harder to collect than the highly 
institutionalized record of actual transactions, to our knowledge 
this investigation was never before attempted by members of the 
econophysics community.  Second, we hope that the study of a 
real order book dynamics would suggest new realistic ingredients that 
can be added to a toy model of Ref [8] to improve its agreement 
with the extended set of stylized facts.  

Markets differ from each other in
precise rules of submission of orders and the transparency of the order book.
In the so-called order-driven markets there are no designated market
makers who are required to post orders (quotes) on both bid and ask sides
of the order book. Instead the liquidity is provided only by 
limit orders submitted by individual investors.  
Versions of this market mechanism are employed in such markets as 
Toronto Stock Exchange (CATS), Paris Bourse (CAC), Tokyo Stock
Exchange, Helsinki Stock Exchange (HETI), Stockholm Stock Exchange
(SAX), Australian Stock Exchange (ASX), Stock Exchange of Hong Kong
(AMS), New Delhi and Bombay Stock Exchanges, etc.
Major US
markets use somewhat different systems. In the New York Stock Exchange 
individual orders are matched by a specialist who does not disclose 
detailed data regarding the contents of 
his order book. That reduces the transparency (or openness)
of the order book to market participants. The NASDAQ Level II screen
is the closest US equivalent to an order book 
in an order-driven market. Since the contents and dynamics 
of individual entries on 
this screen are main subjects of the present work they will be described 
in greater details later on in the manuscript. 

Before we proceed, we would like to put an important disclaimer
regarding the terminology used in this paper. To avoid
overwhelming our readers by a variety of different financial terms
describing similar concepts, in this work, we would refer to 
any yet unfilled order present in an order book as a `limit order.'  
While this is strictly true for an order 
driven market, using this term to describe a market 
maker's quote on the NASDAQ Level II screen may 
seem a bit confusing at first.  However, it makes sense 
in this context.  Indeed, both individual limit orders in an 
order-driven market and market maker's quotes on the 
NASDAQ Level II screen can be viewed just as commitments 
to buy (sell) a certain number of shares at a given price should 
the queuing mechanism match this order with a complement 
marketable order. The only detail which distinguishes a 
market maker from a normal trader in an order-driven market 
is that by NASDAQ rules, the market maker must maintain 
both buy and sell limit orders, changing price level and 
volume within domains established by exacting timing rules.  
But in zero order approximation one can simply forget 
that these two quotes come from the 
same source and look at them just as at 
two individual `limit orders.'  

The other simplification adopted in this work is 
that we do not make a distinction between a true market
order and a marketable limit order, placed at or better than the 
inside bid or ask price, and refer to both of them as
`market orders'. From this point of view a transaction always happens
when a `market order' (or a marketable limit order) is matched with a
previously submitted `limit order' (or a quote by the market maker).
The size of an individual transaction is therefore a good measure of 
a market (or marketable) order size in our definition.

The real time dynamics of an order book is a fascinating       
spectacle to watch (see e.g. www.3dstockcharts.com). 
For frequently traded stocks it is in a state of a constant change. 
%
%
The density of limit orders goes up when more traders 
select to submit limit orders rather than market (or
marketable) orders. 
In the opposite case of a temporary preponderance of 
market orders, the book gets noticeably thinner.
In addition to these fluctuations in the density and number 
of limit orders, any serious imbalance in the number limit orders to
buy and limit orders to sell near the current price level 
gives rise to short term deterministic price changes. 
This change reflects intuitive notions
regarding supply and demand. i.e. the price statistically tends
to go up in response to an excess number of limit orders to buy and
down in the opposite case. It is by observing all of this in real 
time one understands that the balance of individual orders in the
order book is the {\it ultimate} source of price fluctuations. 

In this work we study the statistical properties of data one of us (MM) 
collected on the NASDAQ market.  Even though NASDAQ is a quote-driven (dealership) market,  
due to reasons explained above we believe that our study
should also apply to order books in order-driven markets.  
Indeed, many of our conclusions are remarkably similar to those 
reported for order-driven markets in the
recent economic literature \cite{econo}.
The NASDAQ Level II data for a given stock lists 
current bid and ask prices and volumes quoted by all market makers
and Electronic Communication Networks trading this stock. 
For example the line: 
JDSU    GSCO      K       NAS      112.625     500 114.0625     500 
can be interpreted as a display of Goldman Sachs'(GSCO) intent 
to buy 500 shares of JDS Uniphase Corporation (JDSU) at 112.625 per
share and sell 500 shares at 114.0625 per share. Each such 
market maker entry usually conceals a whole secondary order book of 
limit orders submitted to this market maker by his clients.
Those `outside' bids and asks, i.e. private limit orders at price
levels more distant from the publicly displayed
`best' bid or ask, generally remain hidden to most market
participants.  
The concept of second hierarchical level of order books at NASDAQ 
can be perhaps best illustrated on an example of Electronic
Communication Networks (ECN) such as Island (the ECN symbol ISLD). 
In this case the ``hidden'' book can be actually 
viewed (e.g. at the Island's website (www.island.com)), 
while the only part of this book which is visible at the NASDAQ Level
II screen is the highest bid and lowest ask prices and volumes.
There they are shown as any other market maker entry:
JDSU    ISLD      O       NAS       113.75     200     114     800.

In the course of one trading day we recorded 'snapshots' of the order 
book for one particular stock at time intervals which are 
on average 3 seconds apart. We were unable to account for network
delay between our 'time stamp' and the actual display time (in the
NASDAQ order-matching queue).  
The delay was generally assumed to be less than a second, but it is known
exceed 2 or 3 seconds when high trading volume imposed network delays.
This record was subsequently binned by the price, and aggregate 
volumes at four highest bid prices and lowest ask prices
were kept in the file. 
Due to the discreteness of stock price at NASDAQ several market makers
are likely to put their quotes at exactly the same price. In our file 
we kept only the aggregate volume at a given price, equal to the sum
of individual limit orders (quotes) by several market makers. 
A file collected during a typical trading day contains on average 
7000 time points.

The first question we addressed using this data set was: 
what is the size distribution of limit and market orders? 
In Fig. 1 we show the cumulative distribution of {\it market} (marketable)
order sizes (or alternatively the sizes of individual transactions) 
calculated for all stocks and trading days for which we have 
collected the data. 
\begin{figure}
\centerline{\epsfxsize=5in
\epsffile{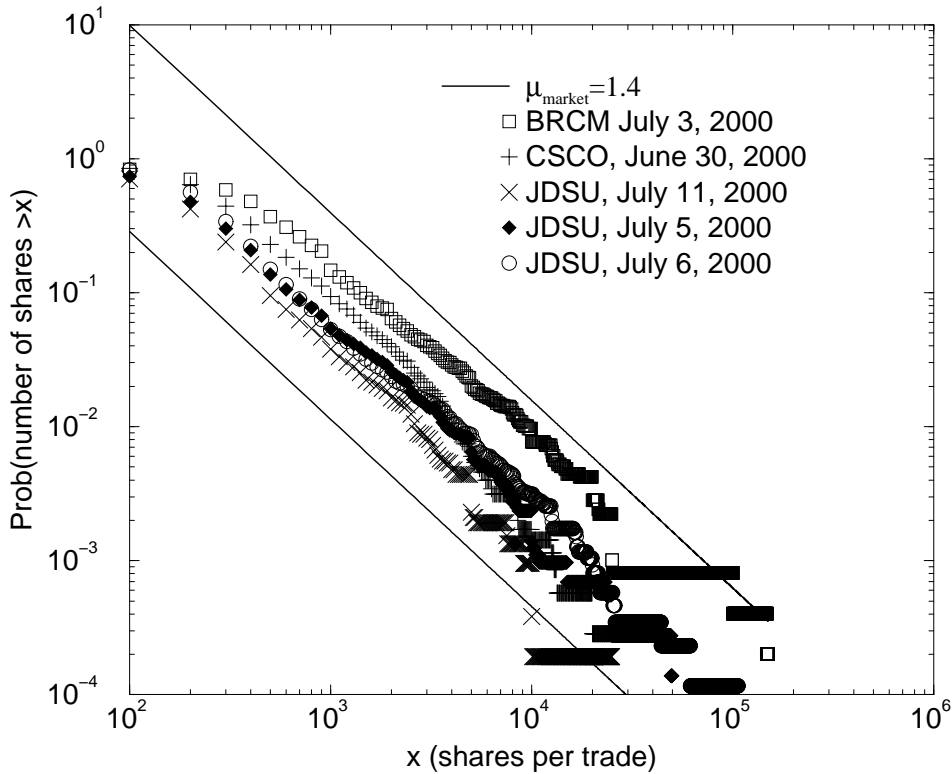}
}
\caption{
The cumulative distribution of market order sizes. The
straight line has the slope $\mu_{market}=1.4$.}
\label{fig1}
\end{figure}   
From our record we know only the total number of traded shares
and the total number of transactions which occurred between the two
subsequent snapshots of the order book.
This average number of transactions per snapshot 
varies between 3 and 5.5 for different stocks in our data set. 
The size of a market order used in Fig. 1 was defined simply as the
change in the traded volume 
divided by a small number of transactions that occurred between the two
subsequent snapshots of the screen. All our data are consistent with 
market order sizes being distributed according to a power law 
$P(x) \sim x^{-1-\mu _{market}}$ with an exponent $\mu_{market}=1.4
\pm 0.1$.  In \cite{gopi2} Gopikrishnan {\it et al.} have 
analyzed the distribution
of volumes of individual transactions for largest 1000 stocks traded 
at major US stock markets and arrived at a similar {\it average} 
value for the  exponent $\mu_{market}=1.53 \pm .07$ ($\xi$ in their
notation). They also plotted the histogram of this exponent measured 
for different individual stocks (see Fig. 3(b) in 
Ref. \cite{gopi2}), showing substantial variations. 
     
The distribution of {\it limit order} sizes, to our knowledge, was
never analyzed in the literature before. To make the
histogram of this distribution we used sizes of limit orders at 
a particular level in the order book from all snapshots made 
throughout one trading day.
We found that this histogram can be also approximately described by a 
power law form. The data for different levels of bid
and ask prices (level 1 being the highest bid and the lowest ask) 
for two of our stocks are presented in Figs. 2,3. 
\begin{figure}
\centerline{\epsfxsize=5in
\epsffile{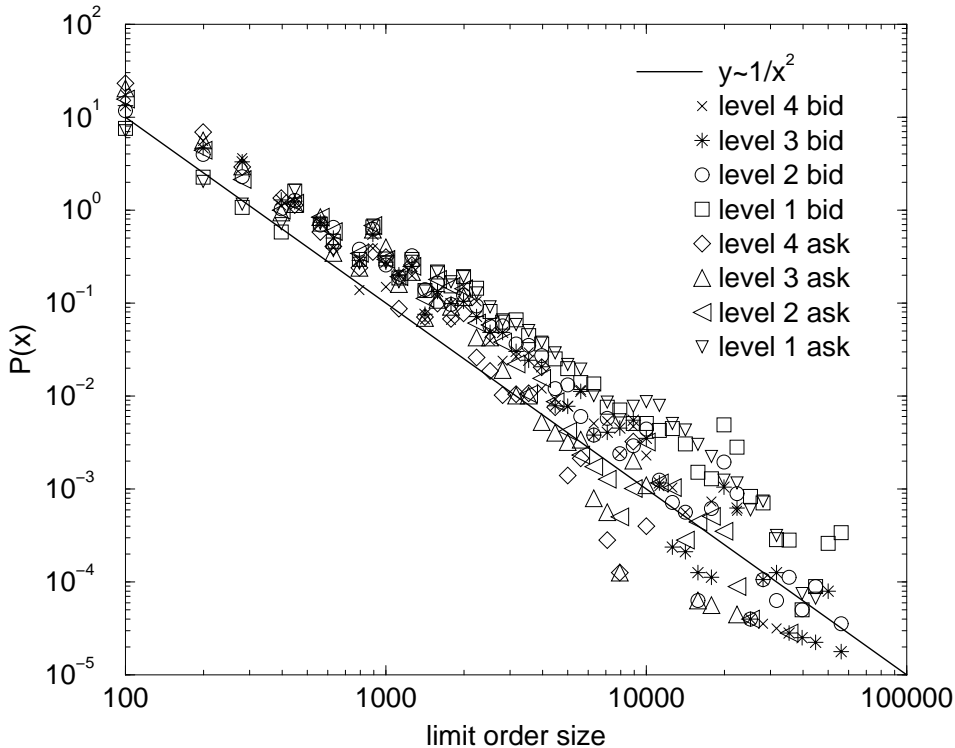}
}
\caption{
The size distribution of limit orders (consolidated 
market maker quotes) for the stock of 
the JDS Uniphase Corporation (ticker symbol JDSU) traded on July 5,
2000. The straight line has the slope $1+\mu_{limit}=2$.}
\label{fig2}
\end{figure}  
\begin{figure}
\centerline{\epsfxsize=5in
\epsffile{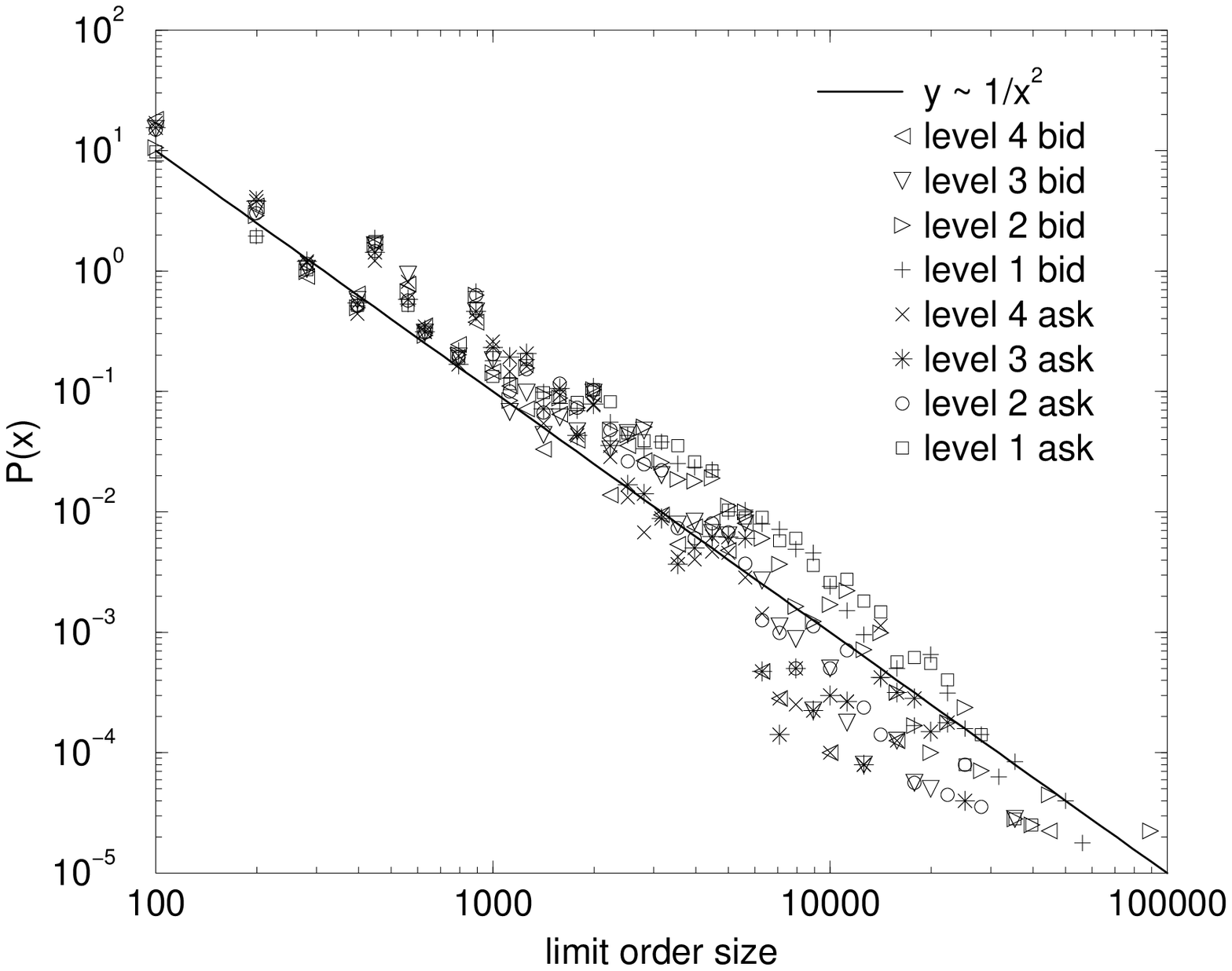}
}
\caption{
The size distribution of limit orders (consolidated market maker quotes) 
for the stock of the Broadcom Corporation (ticker symbol BRCM)
traded on July 3, 2000. The straight line has the slope 
$1+\mu_{limit}=2$.}
\label{fig3}
\end{figure}  
In both cases all distributions were found to be consistent with 
an exponent $\mu_{limit}=1.0 \pm 0.3$. 
The quality of the power law fit is rather poor though.
In fact when we repeated the above analysis using cumulative
histograms we saw that a log-normal distribution fits 
our data over a wider region (see Fig. 4).
The best fit to a log-normal distribution has similar parameters 
for different stocks, trading days, and levels in the order book. 
The best empirical formula for the probability distribution 
of limit order sizes is thus
$P(x)=x^{-1} \exp(-(A-\ln(x))^2/B)$, with parameters $A$ and $B$
fluctuating around 7 and 4 in all of our data sets. This formula 
indeed gives the effective power law exponent $\mu_{limit}=1$ 
for $x \simeq 8000$ i.e. near the center of our range.
\begin{figure}
\centerline{\epsfxsize=5in
\epsffile{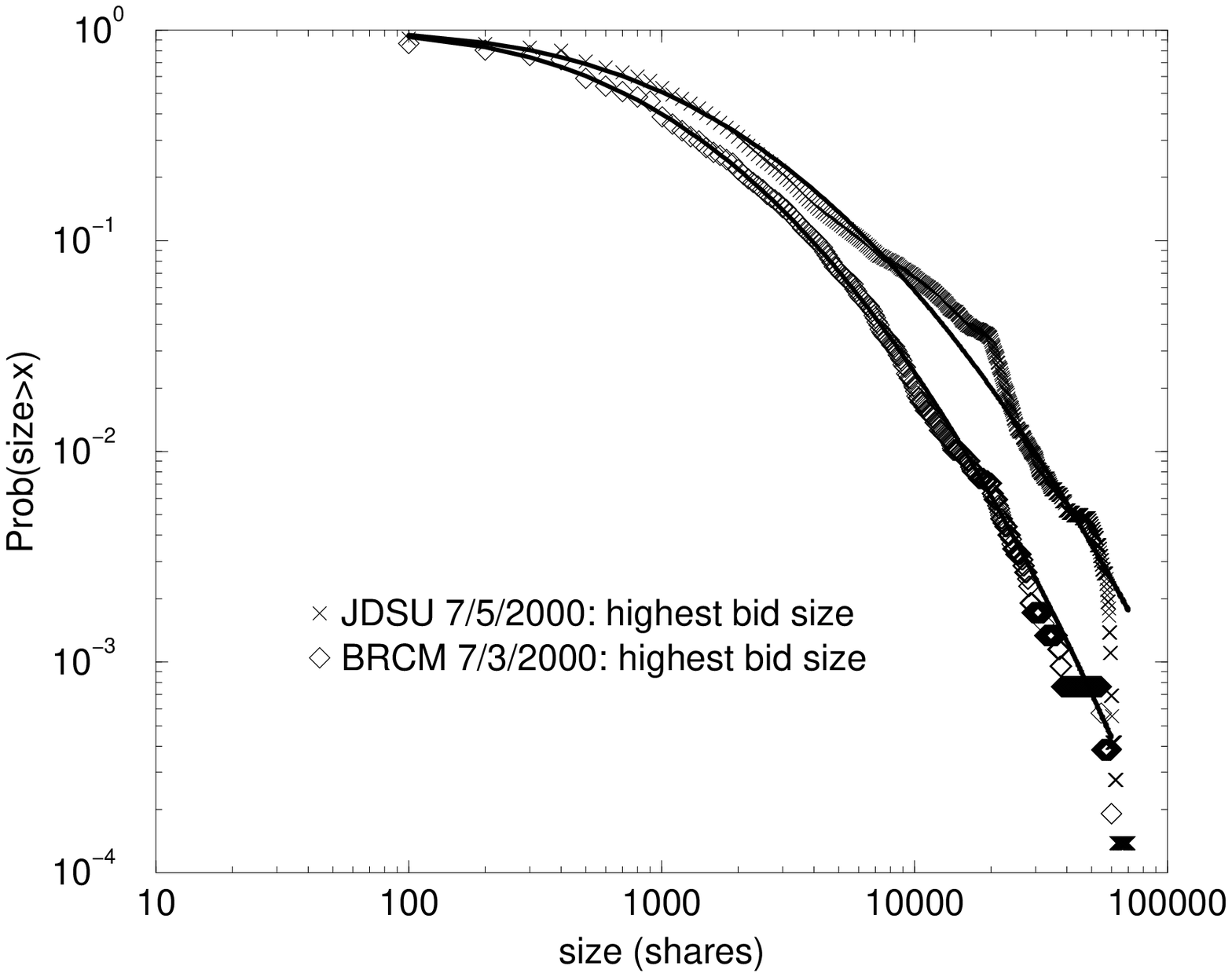}
}
\caption{
The cumulative distribution of highest bid sizes for 
stocks and trading days used in Fig. 2 and 3. 
Solid lines are best fits with the
cumulative histogram of a log-normal distribution 
$P(x)=x^{-1} \exp(-(A-\ln(x))^2/B)$. The best 
fit parameters are $A_1=6.94$
and $B_1=4.20$ for the JDSU and $A_2=6.57$
and $B_2=3.56$ for the BRCM. }
\label{fig4}
\end{figure}
  

We next concentrate on calculating the depth of the order book 
at any given bid and ask level. The depth of the order 
book is an important
measure of the liquidity of the market for a given stock.  For a given
state of the order book one can measure the total volume (number of
shares) $N(\Delta p)$ of limit orders with execution prices lying 
within a certain price range $\Delta p$ from
the middle of the bid/ask spread. The function $\Delta p(N)$, which is 
the functional inverse of $N(\Delta p)$ can be thought of as a 
{\it virtual} impact that a hypothetical market order of volume $N$ would have 
on the price of the stock. It is important to emphasize the word virtual
here. Indeed, in real markets new limit orders would be immediately 
submitted by market makers (or speculators in order-driven markets) 
in response to the arrival of a large market order. 
The first step in quantifying the depth
of the limit order book is to measure the average price difference 
between different levels of the book e.g. the average gap between prices of the 
highest bid and the next highest bid. For both bid and ask sides of 
the book at all levels the average price gap between levels 
was measured to be around \$0.08 for the JDSU stock
traded on July 5, 2000 and \$0.12 for the BRCM stock
traded on July 3, 2000. The average bid-ask spread (i.e. the
difference between the lowest ask and the highest bid prices) was
measured to be some 10-20\% smaller than the average gap between 
two levels on the same side of the book. 
Also in both data sets that we analyzed, gaps on the ask (limit orders to
sell) side seem to be some 5-10\% higher than on the bid 
(limit orders to buy) side. It is not clear if that was just a
trading day artifact or a sign of some real asymmetry.
More interesting behavior was observed for the average size of a limit
order as a function of the level of the order book.  
The average size is at its maximum at the level 1 of the book 
(highest bid/lowest ask) and gradually falls off with the level number 
(see Fig.5). 
\begin{figure}
\centerline{\epsfxsize=5in
\epsffile{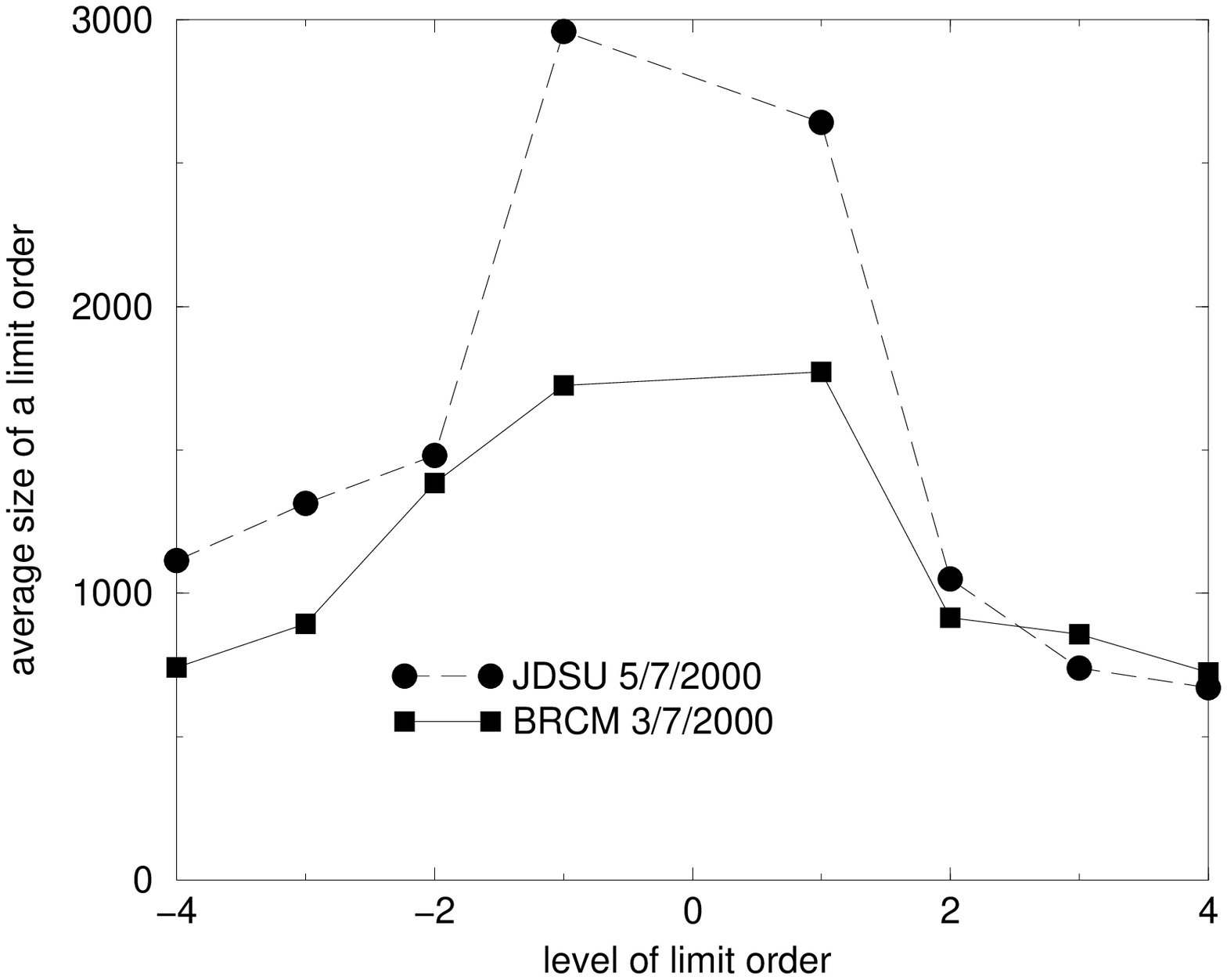}
}
\caption{
The average number of shares offered in limit orders at a given level 
as a function of the level number. Negative levels correspond to 
limit orders to buy (bids), while positive to limit orders to sell 
(asks).}
\label{fig5}
\end{figure} 
Using the
data for the average volume at each level and the average price 
difference between levels one easily reconstructs the average 
virtual impact curve. From Fig. 6 one concludes that the {\it virtual} price
impact $\Delta p(N)$  of a market order is a nonlinear function 
of the order size $N$. Similar results were observed for the limit
order book at the Stockholm Stock Exchange by Niemeyer and Sandas (see
Fig. 8 in \cite{niemeyer}).
\begin{figure}
\centerline{\epsfxsize=5in
\epsffile{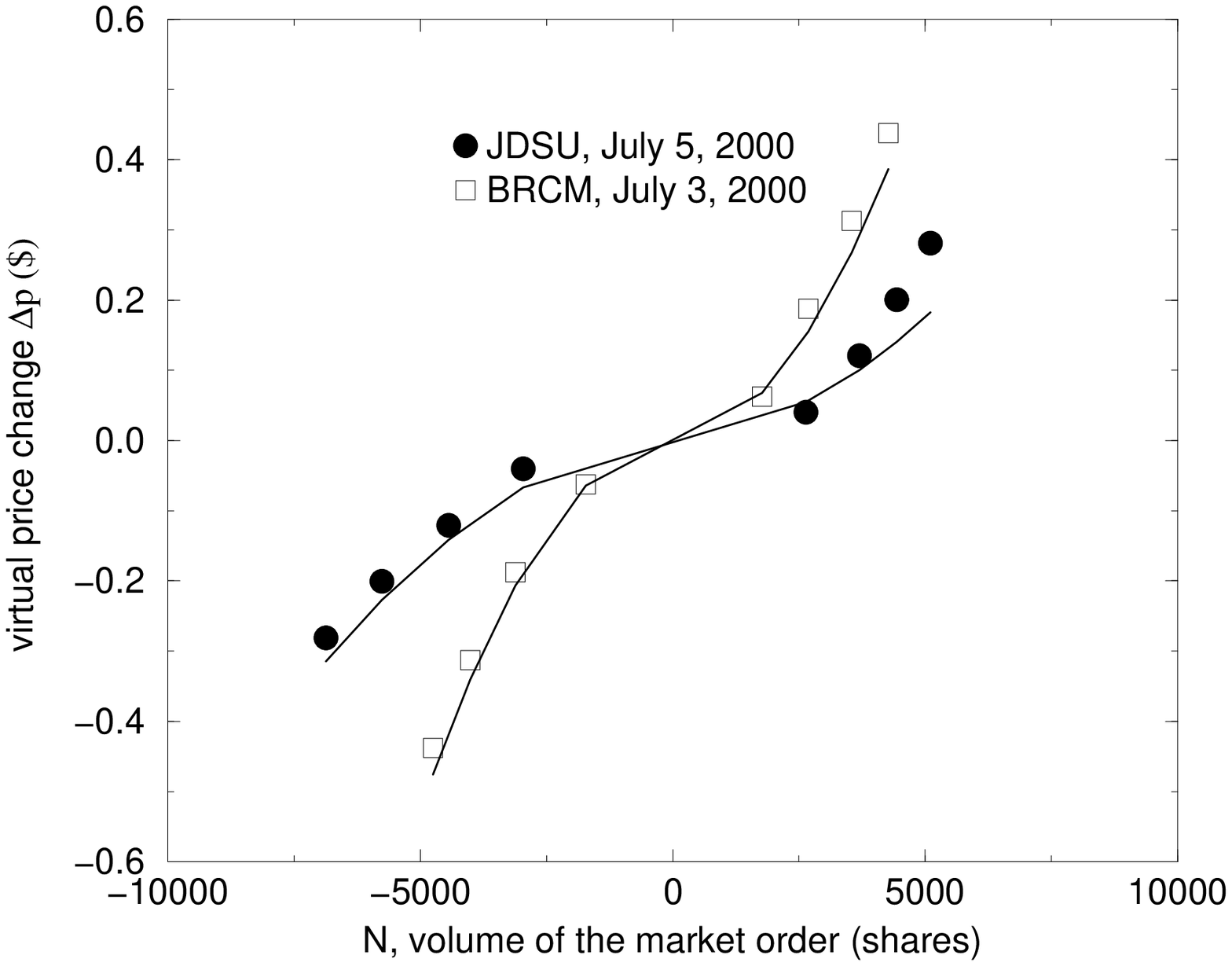}
}
\caption{
The virtual impact of a market order calculated from the
density of limit orders in the order book. 
Negative $x$ corresponds to market orders to sell, 
while positive - to market orders to buy. Solid lines are an attempt
to fit the data with the power law form. The exponent $\delta$ of the
best fit was close to 2 in both cases.}
\label{fig6}
\end{figure}
To have a concise formula for $\Delta p(N)$ we 
fit it to the power law $\Delta p(N) \sim N^{\delta}$.
The exponent $\delta$ in this fit fluctuated
between 1.7 and 2.2 in different data sets. 
In Ref. \cite{manifesto} it was argued that 
the price impact function should have an exponent $\delta=0.5$. 
This conjecture was later used in several models
to arrive at the empirically observed value of the exponent $\alpha$
of the fat tails of the histogram of price fluctuations.
Our virtual market impact function characterized by $\delta \simeq 2$ 
has the opposite convexity compared to that with $\delta=0.5$.  We attribute
this discrepancy to the difference between virtual and real market
impacts, where the latter is dramatically softened by actions of 
speculators. 

The subject of speculators brings us directly to the last question we
addressed using our data: can one use the information contained in the
order book to predict the magnitude and direction of 
price changes in the near future? Many seasoned day traders 
would answer yes to this question. From the law of supply and demand
one expects that a significant excess of limit orders to sell above 
limit orders to buy (excess supply of stock) would push the price down 
while in the opposite case the price would go up. 
It means that a speculator who has
access to the current state of the order book 
can predict (and use this prediction for his/her
profit) the direction of price change in the near future. 
The practical applicability of this strategy is limited by the 
fact that all NASDAQ traders have at least
2 routers between them and the 'order-matching queue.' Each router
introduces a network delay. The induced delay may be greater than
the deterministic correlation timespan.  In this work we made no attempt to see if
deterministic correlations existed in 'real time' while data was being
collected.
The first way to measure the short term predictability of market 
price from our data 
is to concentrate on those moments in time when the 
total number of shares contained in 
limit orders to sell and limit orders to buy
differ by a significant number of shares. In principle this
amount should be selected proportional to the average daily volume of
transactions for each particular stock, yet in our calculations 
we fixed it to be 10000 shares for each of the stocks in our 
data sets.  Also, we looked only at the 
imbalance between volumes offered at highest bid and lowest ask 
prices. For one of our data sets we checked that if higher
levels are included our conclusions remain qualitatively the same. 
We then averaged the evolution of price immediately after the 
moment of large excess demand (or supply) over all events when this 
excess was realized.
In Fig. 7 one can see that indeed as can be expected from the law 
of supply and demand an excess demand drives the price up, 
while an excess supply drives it down. 
In our data set
this predictability of future prices lasts only 
for a few minutes (even for 30 seconds for some of the stocks). 
Therefore, speculators who want to use this effect need to act quickly
and to have a very fast and reliable connection to main computers
at NASDAQ. Yet another way to visualize the effect of the 
imbalance of supply and demand on future prices is to calculate the average 
change in price of the stock during a fixed time interval $\Delta t$ 
conditioned at a certain value of the imbalance of the order book 
before the change.  In Fig.8 we plot the average 1-minute
price change as a function of the initial imbalance of limit orders 
at the highest bid/lowest ask levels. 
This plot once again confirms that the influence of the state of the 
order book on future prices is a real and sizable effect. 
At our level of statistical errors it appears that the price change 
scales approximately linearly with the excess supply (or demand). 

In conclusion, we have presented an empirical study of statistical
properties of a limit order book using the high frequency 
data collected in the NASDAQ Level II system. It was observed 
that the distribution of 
market (or marketable limit) orders has power law tails characterized
by an exponent $1+\mu_{market}=2.4
\pm 0.1$. The distribution of limit order sizes is also consistent 
with a power law with an exponent close to $2$. 
However, it was found that a log-normal distribution gives a better 
fit to the cumulative distribution of limit order sizes over a wider range. 
The depth of the order book measured as a virtual price impact of 
a hypothetical large market order was found to be 
a non-linear function of its size. This non-linearity is primarily due
to the decay in the density of limit orders (quotes) away from the 
most recent transaction price. In reality though this virtual impact 
is probably much softened by actions of speculators, so that the
convexity of the non-linear part may even change its sign. 
A large imbalance in the number of 
limit orders at the highest bid and lowest ask sides of the book leads to the
deterministic price changes which are in accord with intuitive 
notions regarding supply
and demand. This effect seems to disappear at a time scale of 
several minutes. The short-term average price change linearly depends
on the imbalance in the total volume of limit orders
at the inside bid and ask prices. 
These empirical  findings may prove to be useful in narrowing down 
the list of models, used to explain the set of stylized facts about
market price fluctuations. Even more importantly, this work
may shift the attention of the econophysics community towards more realistic 
order book based price formation mechanisms. The work is currently
underway to add some of the observed empirical features to the 
simple toy model of order-driven markets proposed by one of
us in Ref. \cite{maslov}. In particular we plan to check the
effect that broad (power law) distributions of limit and market order
sizes would have on the critical exponents of this model.


\begin{figure}
\centerline{\epsfxsize=5in
\epsffile{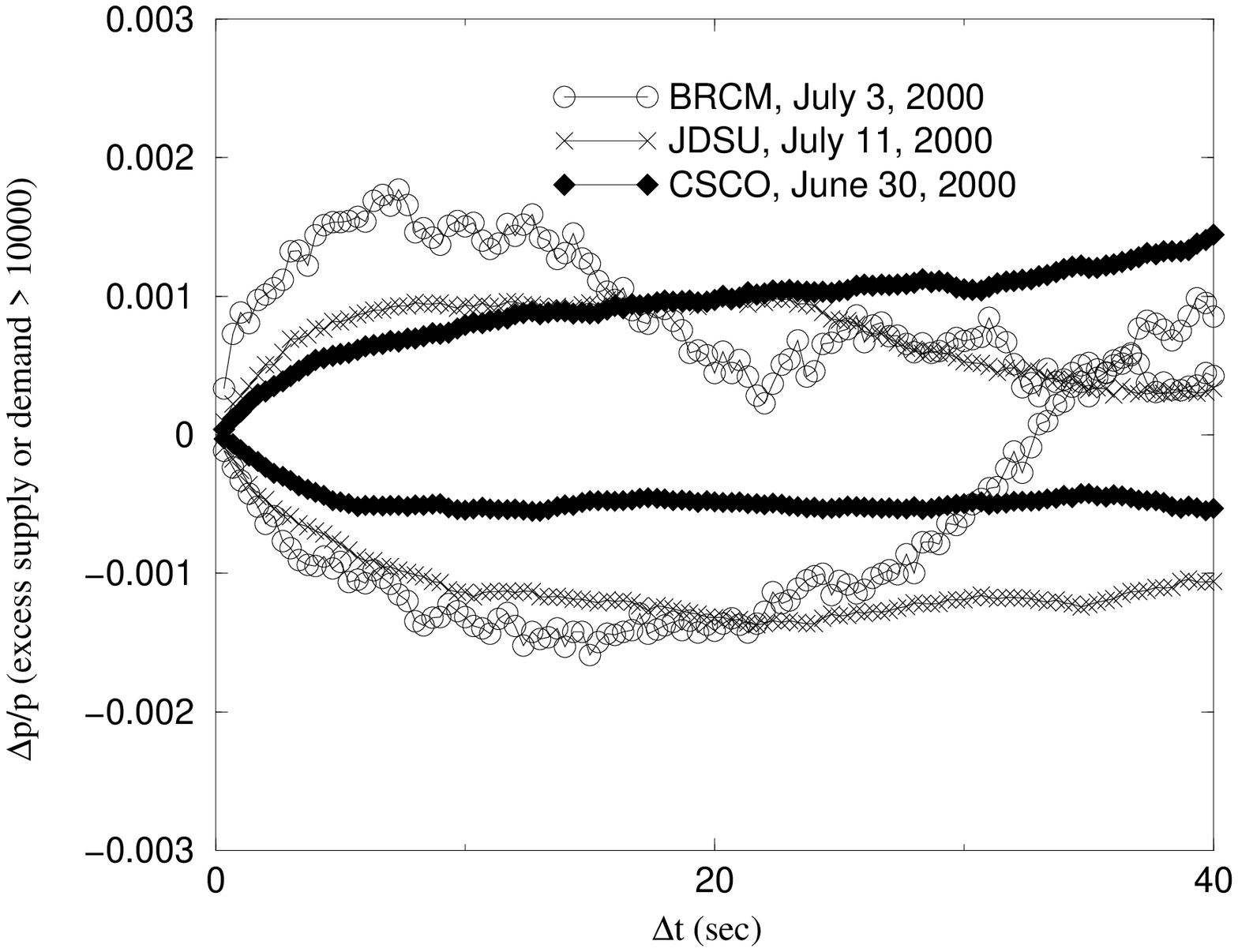}
}
\caption{
The market impact of a large imbalance (larger than 10000 shares)
of the number of shares offered at the highest bid and lowest ask
prices. The upper portions of curves correspond to the excess demand
for the stock, while lower ones for the excess supply. The $y$-axis
shows the normalized price change averaged over all events: 
$\langle (p(t+\Delta t)-p(t)/p(t) \rangle_{t \in Events}$}
\label{fig7}
\end{figure}  

\begin{figure}
\centerline{\epsfxsize=5in
\epsffile{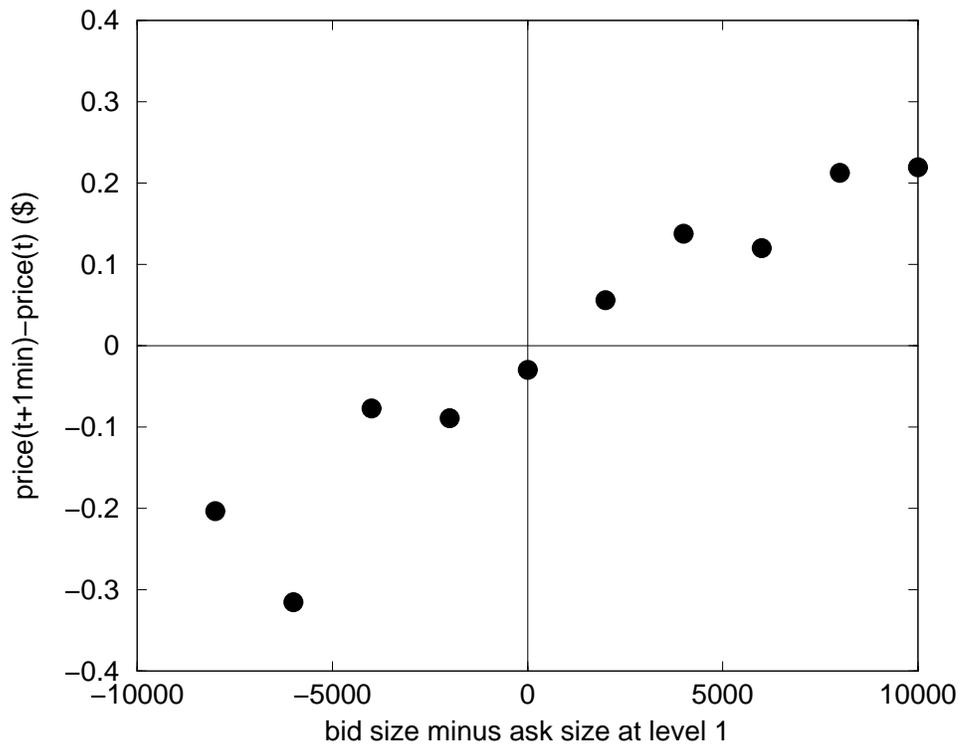}
}
\caption{
The average price change one minute after an excess supply or demand
was observed as a function of the excess demand, i.e. volume at the
lowest ask minus volume at the highest bid.}
\label{fig8}
\end{figure} 

\end{document}